# Optimizing B2B Product Offers with Machine Learning, Mixed Logit, and Nonlinear Programming[1]

John V. Colias, Ph.D., Affiliate Assistant Professor, University of Dallas

Stella Park, Lead Marketing Manager, AT&T Business.

Elizabeth Horn, Ph.D., Senior Vice President, Advanced Analytics Group, Decision Analyst, Inc.

**Abstract**

In B2B markets, value-based pricing and selling has become an important alternative to discounting. This study outlines a modeling method that uses customer data (product offers made to each current or potential customer, features, discounts, and customer purchase decisions) to estimate a mixed logit choice model. The model is estimated via hierarchical Bayes and machine learning, delivering customer-level parameter estimates. Customer-level estimates are input into a nonlinear programming next-offer maximization problem to select optimal features and discount level for customer segments, where segments are based on loyalty and discount elasticity. The mixed logit model is integrated with economic theory (the random utility model), and it predicts both customer perceived value for and response to alternative future sales offers. The methodology can be implemented to support value-based pricing and selling efforts.

Contributions to the literature include: (a) the use of customer-level parameter estimates from a mixed logit model, delivered via a hierarchical Bayes estimation procedure, to support value-based pricing decisions; (b) validation that mixed logit customer-level modeling can deliver strong predictive accuracy, not as high as random forest but comparing favorably; and (c) a nonlinear programming problem that uses customer-level mixed logit estimates to select optimal features and discounts.

**Keywords/Phrases:** B2B, sales data, machine learning, mixed logit, nonlinear programming

# Introduction

In highly competitive Business-to-Business (B2B) markets, sales teams are pressured to increase discounting to retain and win customers. According to Hinterhuber et al., discounts have been a key selling tool in B2B markets, but creating value through product differentiation should be the future focus (Hinterhuber et al. 2021). That being the case, a key challenge is to select optimal product features and price to optimize the tradeoff between market share and profit.

---

Analytically, the optimization problem requires a prediction of offer acceptance where the prediction takes into account the customer's perceived value of product benefits relative to the price paid. The importance of value quantification at the level of sales and account managers within B2B firms has been emphasized in numerous studies (Hinterhuber 2017). Classen and Friedli (2019) explored the literature on value-based marketing and sales (VBMS), and they found 25 articles published between 2015 and 2018 (Classen and Friedli 2019). Most of these studies emphasize a bottoms-up approach to quantifying customer value by identifying the components of value, assigning a value to each component, and using these quantifications in the sales process to convince the potential buyer that the value of the benefits exceeds the price. The method here, instead, uses predictive models to directly measure customer-perceived value within a predictive model. This approach would complement the bottoms-up methods in value quantification by providing a tool to quantify customer-perceived value from an historical offer database and using that quantification to select levels for price and product features to optimize profits.

To predict offer acceptance, one needs an offer database that records offers made by sales teams, those offers that were accepted by the customer, and those which were not. Fortunately, such data is available. While the initial work used real data, that data could not be used in this study due to proprietary concerns. This study used simulated data that mimicked the real data in terms of available variables and their interdependency. The R code used to simulate the data is available upon request. The simulated data includes responses of 1000 customers to product offers, where offers differ across customers and across observations for each individual customer. The offers differ in terms of offer price and contract length. Careful steps were taken to assure that the simulated data represented real data for the purpose of demonstrating analytical process and predictive accuracy.

For proprietary reasons, the real sales offer data could not be used, even if only to assess predictive accuracy. A search for publicly available sales offer data (with 2 years of historical offers at the customer level) yielded no such data. For this reason, e-commerce sales data, publicly available, was used to measure predictive accuracy of the modeling method. While e-commerce sales differ from sales offer data, in that offers are not proactively given to individual customers, the data did resemble the sales offer data in that it included 2 years of historical purchase data at the customer level and included multiple observations per customer.

A hierarchical Bayes estimation of mixed logit model parameters at the customer-level was used to predict offer acceptance. Since choice models such as the mixed logit model are integrated with economic utility theory, they can predict offer acceptance and also measure the value of an offer. In choice models, value in dollars is defined as the utility of the brand and product features divided by an estimated price coefficient (K. Train and Weeks 2005; K. E. Train 2009). In contrast, machine learning models generally have not been integrated with economic utility theory as the prediction goal is the primary emphasis (Zhao et al. 2020).

Key research questions were:

1) Would a typical offer database in a B2B technology market provide sufficient variability of data to estimate a usable mixed logit model?
2) How accurately would a mixed logit model, with customer-level parameter estimates, predict versus benchmark machine learning models to predict customer response to discounted offers?
3) How would one optimize profit using mixed logit model customer-level outputs?

## Literature Review

This study builds on five research areas: (1) choice modeling applications in the support of pricing decisions; (2) value-based pricing and value-based selling in B2B markets; (3) applications of the mixed logit model and hierarchical Bayes to estimate customer-level parameters; (4) investigations into the relative strengths and weaknesses of the mixed logit model and various machine learning models; (5) using nonlinear programming methods to select optimal features and prices.

### Choice Modeling in Support of Pricing Decisions

Many studies have used choice modeling in estimating price effects since the late twentieth century. Examples include a study on the apartment rental market (Elrod et al. 1992), water resources (D. Hensher et al. 2005), camera and automotive (Sonnier et al. 2007), tourism (Masiero et al. 2015), airline industry (Newman et al. 2014; Ratliff et al. 2008; Talluri and Van Ryzin 2004; Wardell et al. 2008), nursing homes (Milte et al. 2018), healthcare (Regier et al. 2009), and transportation (Li et al. 2010).

Classic texts on choice modeling include explanations of how to estimate price effects and (Ben-Akiva and Lerman 1985; K. E. Train 2009). One among many examples of using choice modeling to estimate willingness-to-pay from survey-based stated choices is found in a study of household willingness-to-pay for water service attributes (D. Hensher et al. 2005).

In a general overview of the role of measurement models (Breidert et al. 2006), the authors outline alternative choice modeling approaches including those that use real market data and survey data. The authors point out that real market data has the benefit of face validity but concludes that price variations are normally very limited making their use for pricing decisions problematic.

While that may be true in many applications, this research uses historical offer data in a B2B market where sales representatives customize features and price at the individual customer level via discounting, creating considerable price variation.

The theoretical foundation for choice modeling is the random utility model derived from the field of economics (Ben-Akiva and Lerman 1985). Utility $U_j$ derived from the use of a product $j$ is comprised of an observed utility $V_j$ and an unobserved utility $e_j$.

$$U_j = V_j + e_j$$

If $V_j = k_j + x_j\beta + p_j\gamma$

where

$k_j$ = a scalar constant term for alternative j

$x_j$ = a $1 \times K$ vector of observed features of alternative j, excluding price $p_j$

$\beta = K \times 1$ vector of parameters

$p_j$ = price of alternative j

$\gamma$ = a price parameter measuring price sensitivity

then willingness-to-pay = $(k_j + x_j\beta)/\gamma$ (K. Train and Weeks 2005).

**Value-Based Pricing and Value-Based Selling in B2B Markets**

The importance of measuring in B2B markets rests upon the literature stream on value-based pricing (Hinterhuber 2004, 2008; Wardell et al. 2008; Wu et al. 2019), value-based selling (Töytäri and Rajala 2015), and customer value quantification (Hinterhuber 2017).

Based on Toytari and Rajala (2015, p102), customer value is the "difference between the perceived benefits received and the perceived sacrifices made by a customer". Value-based selling is "a sales approach that builds on identification, quantification, communication, and verification of customer value" (Toytari and Rajala 2015, p.101).

Hinterhuber (2017) points out that, to realize the benefit from value-based pricing and value-based selling, the quantification of value is essential. The ability to quantify value becomes even more crucial as premium pricing decisions are beginning to be made by machine algorithms (Hinterhuber and Liozu 2018). In a literature review on value-based marketing for industrial smart services, data-driven services that use digital technology, the authors conclude that more research is needed in analytic methods to quantify value and develop value-based prices (Classen and Friedli 2019).

Some studies have focused on the use of value calculators that identify and assign a value to each unique component of a product offering (Pöyry et al. 2021). The present study instead uses a choice modeling method to measure willingness-to-pay, used as a basis for selecting optimal product features prices.

## Mixed Logit Model and Hierarchical Bayes to Estimate Customer-Level Parameters

The next research stream used in the study is that of the mixed logit model (D. A. Hensher and Greene 2003; McFadden and Train 2000) and the hierarchical Bayes estimation method which enables estimation of customer-level parameters and thus measure price response and willingness-to-pay at the individual customer level (Rossi et al. 2012; K. Train 2001).

McFadden and Train (2000) show that the mixed logit model, with a fully-specified multivariate covariance matrix for estimated utility coefficients, can approximate any choice model with any distribution of preferences. The mixed logit probability of purchase is:

$$Probability_{ci}(\beta_c) = \frac{e^{u_{ci}}}{\sum_{j \in \{\text{alternatives}\}} u_{cj}}$$ where $u_{ci}$ = utility that customer c assigns to alternative i

$\beta_c$ is a random parameter with a distribution across the population of customers (current or potential).

Utility is a linear function of observed attributes plus an error term $\epsilon$ distributed as iid (independent and identically distributed) extreme value:

$$u_{cj} = k_{cj} + x_{cj}\beta_c + p_{cj}\gamma_c + \epsilon_{cj}$$

where

$k_{cj}$ = a scalar constant term for alternative j for customer c

$x_{cj}$ = a $1 \times K$ vector of observed features of alternative j for customer c, excluding price $p_j$

$\beta_c$ = $K \times 1$ vector of parameters for customer c

$p_{cj}$ = price of alternative j for customer c

$\gamma_c$ = a price parameter measuring price sensitivity of customer c

Note that the subscript c appears in the parameters $k_{cj}$, $\beta_c$, and $\gamma_c$, indicating that mixed logit with hierarchical Bayes estimation delivers customer-level parameter estimates.

## Relative Strengths and Weaknesses of Mixed Logit and Machine Learning Models

One advantage cited for the mixed logit model versus machine learning models (e.g. random forest and gradient boosting models) is that the former is grounded in the random utility model whereas machine learning models are not grounded in any particular economic theory of demand (Aboutaleb et al. 2021; Zhao et al. 2020). For this reason, efforts have begun to compare multinomial and mixed logit versus machine learning models, with respect to predictive accuracy and reasonableness of results.

In an airline application, the latent-class multinomial logit model (able to deliver customer-level estimates by applying segment-level posterior probabilities) was found to predict more accurately than simple multinomial logit, although still less accurately than random forest (Lhéritier et al. 2019). In the Zhao et al. study, the mixed logit model was compared to several machine learning models, and the latter class of models produced greater predictive accuracy, but at the cost of producing reasonable behavioral insights.

In a new stream of research, effort is placed into integrating the theory-based structure of choice models with machine learning models (Van Cranenburgh et al. 2021). In one integration study, a deep learning-based choice model was developed in an effort to improve the predictive accuracy of the MNL while retaining the desired properties of the random utility model (Acuna-Agost et al. 2021). Aboutaleb et al. (2021) estimate a mixed logit model and a nested logit model, where machine learning is used to maximize fit to the data subject to constraints on interpretability. In a study of airport choice, the authors introduce a policy gradient reinforcement learning (PGRL) model. The PGRL model accuracy and interpretability results were found to be superior to those from multinomial logit and random forest (Lu et al. 2021).

The present study does not integrate the two, choice modeling and machine learning. This study, however, examines the achieved accuracy in predicting product offer acceptance for the mixed logit model versus a particular machine learning model, random forest.

One particular advantage provided by the hierarchical Bayes estimation method is explored: estimation of customer-level parameters, one for a product feature (contract length) and one for price.

In the mixed logit model, the multivariate distribution of $\beta_c$ assumes that $\beta$'s may be correlated. Hierarchical Bayes estimation uses estimates of the correlation matrix of the $\beta$'s to learn information about product preferences of customers with more observations available (i.e. customers who have received sales offers more frequently) and to use that learning to more accurately estimate $\beta$'s for customers who have fewer observations, in this case, customers who have received fewer sales offers (Johnson 2000).

As in the Zhao et al. study, the mixed logit model estimates customer-level price effects from training data to predict choices in test data. Zhao et al. (2020) appeared not to use customer-level parameter estimates when predicting test data customer-level choices but used population mean estimates to predict customer-level choices (this detail was not discussed in their article).

Of course, customer-level parameters can be used only to predict response to a next offer given to an existing customer. These parameters do not exist for new customers, thus cannot be used to predict responses to offers given to new customers. For new customers, the mixed logit estimates of population means of parameters could be used to predict at the customer-level.

**Nonlinear Programming to Select Optimal Features and Prices**

Applications of nonlinear programming to select optimal features and prices from choice models is not new. Several recent applications use choice model parameter estimates as inputs to an optimization problem.

One study compared multinomial logit and gradient boosting models. The models were estimated from purchase history data and then input into an optimization problem to determine a display assortment that maximizes profits (Feldman et al. 2018). It was found that the multinomial logit model generated significantly higher revenue per customer visit than the machine-learning method in actual marketing programs, even though multinomial logit predicted less accurately in test data.

A study on parking services used mixed-logit utility functions directly in a mixed integer linear optimization algorithm to support pricing and capacity decisions for parking services (Paneque et al. 2021). In a hotel industry study, the mixed logit model was used to parameterize an optimization algorithm to determine optimal room charge and over-booking level to maximize sales (Saito et al. 2019).

The present study uses mixed logit model parameter estimates at the customer level in the objective function of the optimization problem. More specifically, the optimization problem maximizes profits for the next offer given to each existing customer.

Next-offer profit (NOP) was defined as the present value of profits over the lifetime of the contract period, summed over all next-offer contracts purchased by customers. NOP differs from customer lifetime value (CLV) in that not all future years of profit are included for every customer. With contract-based sales where contract length is negotiable, as in the present study, the present value of next-two-year profits is included if the customer purchases with a two-year contract, but the present value of next-five-year profits is included if the customer purchases a with a five-year contract. Thus, the optimization problem maximizes the present value of future profits captured by the next set of sales offers.

A CLV objective function would be more complete as it would include all future sales efforts, not just the next sales efforts. Fader and Hardie (2016) discuss that the potential error in CLV calculations is large when one assumes a constant retention rate over time (Fader and Hardie 2016), as is often done. On the other hand, there is great uncertainty and error in predicting purchase for sales offers to be made 2, 3, or 4 years in the future. Market conditions can change rapidly, and customer perceptions change over time in uncertain ways. Selecting product features and discount levels to maximize CLV would include much more uncertainty and error in the process, perhaps leading to poor decisions.

The present methodology does not depend on using NOP. CLV could be used. For simplicity in the present study, NOP is used.

## Contribution to the Literature

Contributions to the literature include:

(1) A methodology to measure customer-perceived value for offers made by sales representatives in a B2B market. The literature review of methods used to measure customer-perceived value for the purpose of supporting value-based pricing and value-based selling did not reveal any studies that used the mixed logit model to measure customer value.

(2) The comparison of choice modeling and machine learning methods. Van Cranenburgh et al. (2021) reviewed the literature and found 25 studies comparing choice models to machine learning models. The multinomial logit model (MNL) is used as the benchmark for comparison in 17 studies, and mixed logit is used as a benchmark in only 2 studies. Commenting on this fact, the authors state that "using the MNL model as a benchmark is especially inexpedient for studies with a comparison objective. Since state-of-practice discrete choice models, such as the Mixed logit model, usually outperform MNL models…it is actually unclear how much is really gained by using machine learning models in terms of performance and prediction accuracy, relative to state-of-the-practice discrete choice models".

The present study compares predictive accuracy for mixed logit (using Bayesian estimates of customer-level parameters) versus a machine learning model. The comparison of predictive accuracy for mixed logit using customer-level parameter estimates, has not appeared before in the journal literature.

(3) A methodology to incorporate customer-level parameters from the mixed logit model into a nonlinear mixed-integer programming algorithm to optimize features and price. The methodology provides a data-driven tool to add to existing tools (e.g., value calculators) to assist sellers in B2B firms to successfully implement value-based selling.

## Methodology

The foundation of the methodology is an offer database that records information supplied by sales representatives. The information includes qualitative assessment of customer loyalty and details of every product offer made to customers. The integration of such an offer database into the sales process:

- Forces sales representatives to record information that otherwise would exist only within their own minds
- Makes the information available for optimizing future offers
- Increases customer retention, loyalty, and acquisition if offers are optimized properly.
- Provides a competitive advantage if the information within the sales database is properly leveraged, because competitors do not have access to the same information.

As represented in Figure 1, the methodology for optimizing offers includes customer-level choice modeling to estimate probability of offer acceptance, model training to increase predictive accuracy, simulation to define loyalty segments, and non-linear mixed integer programming to optimize future offers to each loyalty segment.

**Figure 1**

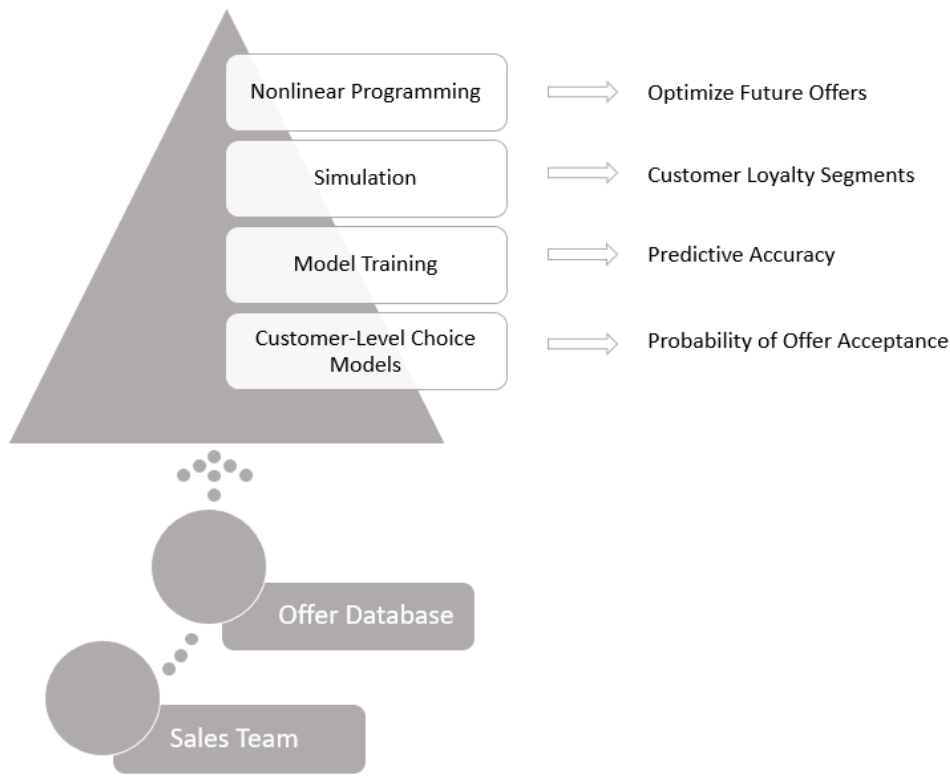

Customer-level modeling is achieved using hierarchical Bayes (HB) choice model estimation. Model training is accomplished using cross-validation resampling to tune hyperparameters of the HB choice model.

**Rationale for Methods Used**

An offer database that contains variability in terms of price and contract length enables development of a predictive model which, in turn, can be used to optimize offers. Several challenges, however, present themselves in the analysis of such data, including

- Accurately predicting the response to a new offer at the customer level. It may be challenging to develop a model for customers to whom very few offers have been made.
- Accurately adjusting predictions if purchased offers are under-sampled relative to non-purchased offers. This may occur if sales reps are responsible for entering offers into the database.

- Optimizing product features and price in a way that accounts for and rewards customer loyalty.
- Incorporating constraints on business actions, for example, capping prices offered to loyal customers.

HB modeling is not commonly used in machine learning. For example, neither the scikit-learn package in Python (Pedregosa et al. 2011), nor the caret package in R (Kuhn et al. 2020b) include HB modeling as defined by Rossi, et al. (2012). This method was adopted here to take advantage of its statistical properties that enable estimation of customer-level model coefficients for all customers, even with sparse data for some customers (i.e., not many product offers), achieved through "borrowing" of information from those customers with more data to better estimate parameters for those with less data (Johnson 2000).

Nonlinear programming was selected because it easily can be modified to incorporate business objectives and constraints that are different from those used in this case study. In other words, it provides a flexible framework adaptable for different business objectives and can incorporate practical considerations such capping prices, maintaining acceptable unit sales, and keeping loyal customers happy by offer price and features that they are willing to pay for.

Commonly used machine learning software available in the R language package, caret, was used to tune the HB model for maximum predictive accuracy.

Finally, open-source software was elected to implement all of the analysis, to demonstrate that this kind of analysis can be implemented without purchasing expensive commercial software. R, the R packages bayesm (Rossi 2017) and caret (Kuhn 2019; Kuhn et al. 2020a, 2020b), Python, the Python package pyomo (Hart et al. 2017) and Bonmin (Pietro Belotti 2019) all were used. (Our R and Python code is available upon request.)

**Hierarchical Bayes Estimation of Customer-Level Parameters of the Mixed Logit Model**

The use of choice modeling grew exponentially during the late $20^{th}$ century, becoming a standard for high-quality marketing research, especially in the area of optimization of product, product line, and price (Agarwal et al. 2015; Wittink and Cattin 1989). The theoretical foundation for choice modeling is the random utility model derived from the field of economics (Ben-Akiva and Lerman 1985). Numerous advances in model specification and estimation have included Nested Logit (Ben-Akiva and Lerman 1985), Mixed Logit (McFadden and Train 2000), the Generalized Random Utility Model (Walker and Ben-Akiva 2002), and model estimation algorithms such as Latent Class (Kamakura and Russell 1989), Simulated Maximum Likelihood (Gourieroux and Monfort 1993), and hierarchical Bayes (Rossi 2017; K. E. Train 2009).

This study uses hierarchical Bayes and the mixed logit model (McFadden and Train 2000) to estimate the probability of purchasing a subscription service as a function of discount and contract length.

The $Probability_{c,o}$ is parameterized as follows, where the offer o is equivalent to the alternative, denoted by i or j in the following formulation.

The mixed logit likelihood function model is:

$$Probability_{ci}(\beta_c) = \frac{e^{u_{ci}}}{\sum_{j\in\{\text{alternatives}\}} u_{cj}}$$ where $u_{ci}$ = utility that customer c assigns to alternative i

$\beta_c$ is a random parameter with a distribution across the population of customers (current or potential).

McFadden and Train (2000) show that the mixed logit model, with a fully-specified multivariate covariance matrix for estimated utility coefficients, can approximate any choice model with any distribution of preferences. Taking advantage of this property, the present model specifies the multivariate distribution of parameters as a mixture-of-normals (Rossi et al. 2012). Machine learning is used to select the number of normal distributions (a hyperparameter).

Utility $u_{cj}$ is a linear function of observed attributes plus an error term $\epsilon$ distributed as iid (independent and identically distributed) extreme value:

$$u_{cj} = k_j + x_{cj}\beta_c + \epsilon_{cj}$$

where

$k_{cj}$ = a scalar constant term for alternative j for customer c

$x_{cj}$ = a $1 \times K$ vector of observed attributes of alternative j for customer c

In this case, two attributes were included, namely, discount percent and contract term.

$\beta_c = \sigma_c \alpha_c$

$\sigma_c$ is a scalar for customer c, inversely proportional to the standard deviation of the error term (Ben-Akiva and Lerman 1985)

In this case, $\sigma_c = 1$, by assumption.

$\alpha_c$ is a $K \times 1$ vector of parameters for customer c

The random error term for customer c, alternative j, quantifies the impact of unobserved factors on the purchase of the particular product by the particular customer. $\epsilon_{cj}$ is modeled as a random term since:

- Unobserved factors can vary over time.
- Which unobserved factors are considered by each customer can change over purchase occasions.

- For the business customer as in the analysis for this paper, unforeseen changes in company policy, job roles, and budgets impact purchase decisions.

In the mixed logit model, the multivariate distribution of $\beta_c$ may be correlated, and the correlation matrix can be empirically measured, given the availability of multiple observations, for example, time-series data, for each customer.

Correlations across the various customers in terms of how much relative weight $\alpha_{ck}$ is placed on each particular observed product attribute $k$, where $\alpha_c = \begin{pmatrix} \alpha_{c1} \\ \vdots \\ \alpha_{cK} \end{pmatrix}$, provides insight into how customers cluster in terms of purchase-decision behavior.

The customer's scale parameter $\sigma_c$ provides insight into the extent to which observable product attributes vs. unobservable factors impact the tendency of the customer to purchase a particular product. However, as aforementioned, $\sigma_c$ is not estimated separately. Rather, the joint parameters $\beta_c = \sigma_c \alpha_c$, are estimated, assuming $\sigma_c=1$.

A hierarchical Bayes method is used to estimate model parameters, because this method

- Is computationally easy to solve (Rossi et al. 2012)
- Provides individual customer coefficients (Huber and Train 2001; Rossi et al. 2012; K. E. Train 2009)
- For customers with little historical data, automatically shrinks customer coefficients towards market averages for similar customers (Rossi et al. 2012).
- Uses correlations of product feature and price preferences across customers to further boost predictive accuracy (Rossi 2017).

It should be noted that, under "fairly benign conditions, the Bayesian estimator is asymptotically equivalent to the maximum likelihood estimator" (Hess and Train 2017).

**Validation of Predictive Accuracy**

As explained in the data section of this paper, actual data (not simulated data) was used to measure the relative predictive accuracy of the hierarchical Bayes mixed logit model vs. a random forest model.

First, a hierarchical Bayes mixed logit model with a multivariate mixture-of-normals distribution was estimated using the R bayesm package (Rossi 2017), explained in detail in chapter 5 of Rossi, et al. (Rossi et al. 2012). The number of components (a hyperparameter) in the mixture of normals were tuned using 5 repeats of random splitting into training and validation data, using the R caret package (Kuhn et al. 2020a) (R code available upon request).

Second, a random forest model (Liaw and Wiener 2002) was estimated, tuning the number of randomly drawn variables (a hyperparameter) included in each classification tree of the ensemble of trees, again using 5 repeats of random splitting into training and validation data.

The predictive accuracy of the two modeling methods was compared using holdout-validation accuracy statistics. In addition, each model was re-estimated with a final training set using the optimal hyperparameters from model tuning and applied to a hold-out sample, and the final predictive accuracy was compared.

Random forest was used as a benchmark to assess predictive accuracy of the hierarchical Bayes model. Random forest is easy to train and tune and produces similar predictive accuracy when compared with other highly accurate models, such as boosting and neural network models (Hastie et al. 2009).

A recent study (Zhao et al. 2020) compared mixed logit to several machine learning models, including random forest, and found the latter to produce greater predictive accuracy, but at the cost of producing reasonable behavioral insights. Their study chose to implement cross-validation resampling with the unit of observation defined as the choice set, rather than the individual. The same approach was selected here. However, the mixed logit model in their study used normal distributions vs. the mixture-of-normal distributions presented here.

Zhao, et. al. (2020) appeared not to use customer-level parameters when predicting choices at the customer level, although this detail was not discussed in their article.

**Nonlinear Mixed-Integer Programming**
Using customer-level predicted purchase probabilities (for the simulated hold-out test data), based on the mixed logit model and estimated with the R bayesm package, a next-offer profit (NOP) objective function then was specified. The NOP objective was maximized given constraints on discount offered within four loyalty/discount elasticity segments. The outcome is an optimized business pricing strategy by customer loyalty/discount elasticity segment.

More specifically, NOP was optimized by selecting a profit-maximizing discount level and contract term by customer segment, where the segments were defined based on discount elasticity and customer loyalty groups.

Discount elasticity groups were based on discount elasticity, $elasticity_c$, in the following way:

If $elasticity_c >= -1$, then $discount\ inelastic$; if $elasticity_c < -1$, then $discount\ elastic$.

Loyalty groups were defined based on $Loyalty_c$:

If $Loyalty_c > 0.5$, then $loyal$; if $Loyalty_c <= 0.5$, then $not\ loyal$.

Discount elasticity, $elasticity_c$, was quantified by using the customer-level models (based on the optimal hierarchical Bayes logit model) to simulate probability of acceptance.

The probability of accepting the offer was simulated for each customer at the discount offered (in the test data); then, an alternative simulation provided acceptance probability with an additional 10% discount. Using an arc elasticity formula, a discount elasticity was estimated for each customer. The arc elasticity is defined as follows:

$$elasticity = \frac{(\overline{Prob_1} - \overline{Prob_0}) \Big/ \frac{(\overline{Prob_0} + \overline{Prob_1})}{2}}{(\overline{Price_1} - \overline{Price_0}) \Big/ \frac{(\overline{Price_0} + \overline{Price_1})}{2}}$$

Table 1 reports the distribution of four segments, based on discount elasticity and loyalty.

**Table 1**

| Discount-elasticity | Loyalty | Percent of Customers |
|---|---|---|
| **Inelastic (elasticity > -1)** | Not Loyal (loyalty score ≤ .5) | 26.5 |
| **Inelastic (elasticity > -1)** | Loyal (loyalty score > 0.5) | 34.8 |
| **Elastic (elasticity ≤ -1)** | Not Loyal (loyalty score ≤ .5) | 22.2 |
| **Elastic (elasticity ≤ -1)** | Loyal (loyalty score > 0.5) | 16.5 |

For the nonlinear programming problem, NOP is maximized subject to upper- and lower-bound constraints on discounts offered within customer loyalty segment.

Select $r_{e,l}$ and $M_{e,l}$ to maximize $\sum_{l \in L} \sum_{c \in C_l} \sum_{o \in O_c} NOP_{c,o}$

subject to

$blower_{e,l} \leq r_{e,l} \leq bupper_{e,l}$

$NOP_{c,o} = (Probability_{c,o}) \times Loyalty_c \times (Present\ Value_{c,o} - Initial\ Cost_o)$

$$Present\ Value_{c,o} = \sum_{m=1}^{M_{e,l}} \frac{MRP_{c,o}(1+r_{e,l}) - MRC_{c,o}}{(1+d/12)^m}$$

$$\forall\ e \in \{discount\ inelastic, discount\ elastic\}, l \in \{not\ loyal, loyal\}$$

where

$Probability_{c,o}$ is the customer-level simulated probability of the purchase of offer o for customer c, based on the choice model

$Loyalty_c$ = customer c's loyalty (a score bounded by 0 and 1). Loyalty is defined as the percent of offers where the sales representative indicated that a strong relationship was the most influential reason for acceptance of the offer. This definition of loyalty was derived from actual sales offer data, which the simulated offer data was designed to mimic.

$Present\ Value_{c,o}$ = the present value of future monthly revenue for customer c and offer o

$Initial\ Cost_o$ = sales and marketing costs for offer o (for simplicity, assumed to be zero in this paper)

$r_{e,l}$ = the discount rate offered for customers of elasticity segment e and loyalty group l

$d$ = the annual rate-of-return for calculating the present value of a stream of future cash flows

$M_{e,l}$ = the number of months, i.e., term of the offer contract for segment e and loyalty group l

$MRP_{c,o}$ = undiscounted monthly recurring price for customer c, offer o

$MRC_{c,o}$ = monthly recurring cost to service for customer c, offer o; A simple $5 per unit monthly recurring cost in the optimization is included in order to demonstrate how costs can be incorporated into the objective function.

## Data

Two datasets were used in this study:

(1) Simulated (fabricated) data was used to demonstrate the analytic methodology and the kind of outputs that can be expected from real data. Careful attention was given to ensure that the simulated data approximated realistic relationships among offer variables and customer responses (purchase, no purchase).
(2) Actual data was used to assess the predictive accuracy of the predictive modeling method used in this study.

The simulated data was based on actual historical sales offer data in a technology sector where the company sells hundreds of products in the business-to-business marketplace, and each customer purchases many products.

The actual sales offer data contained from 1 to 48 observations of offers for each of 717 customers, where observations are product offers to a single customer over a period of 2 years. The variables included:

- The dependent variable, purchase or no purchase of the offer (1 if accepted, 0 if not).
- Contract length (1, 12, 24, 36, and 60 months).
- Product discount offered to the customer (applied to the undiscounted monthly recurring price, unit price × number of units) – varied from -0.5 to +0.5 where the latter represents premium pricing.
- Loyalty was defined as the percent of offers where the sales representative indicated that a strong relationship was the most influential reason for acceptance of the offer.

The simulated data was generated using the following procedure:
1. Actual sales offer data was used to estimate a hierarchical Bayes (HB) mixed logit model with the dependent variable being a binary choice (accept or do not accept the sales offer) and independent variables being a constant term, contract length, and product discount. The HB mixed logit model delivered customer-level coefficients that have a multivariate normal distribution. The loyalty variable was included as a predictor of the means of the multivariate normal distribution (Orme and Howell 2009).
2. A supplemental regression was estimated for each customer-level coefficient as a function of the loyalty variable. Based on p-values (essentially zero) for all loyalty coefficients, that the null hypothesis that the loyalty coefficient is zero was rejected. This suggests that the sales representative's perception of loyalty on the part of the customer should be used to assist in measuring customer behavior.
3. For proprietary reasons, none of the outputs from steps 1 and 2 were directly used in simulation of fabricated data. To accomplish this, a small amount of random variation was added to each coefficient mean from step 1 and each loyalty coefficient from step 2 (influencers of coefficient means).
4. Covariances of model coefficients were approximated using customer-level coefficient estimates from step 1. Since the coefficient distributions from step 1 were not necessarily normally distributed due to (a) inclusion of the loyalty variable and (b) imposition of a negativity constraint on the discount coefficient, model coefficient distributions were approximated with a mixture of 3 normal distributions. Specifically used is a Gibbs sampler as outlined in pages 79-81 of the text by Rossi, Allenby, and McCulloch (Rossi et al. 2012). The rnmixGibbs function from the bayesm package was used for estimation of the moments of the mixture of normal distributions, using 5000 draws in the MCMC chain and a burn-in of 500 draws (Rossi 2017).
5. The randomly perturbed coefficient means and loyalty coefficients (step 3) and the covariance matrices for the 3 multivariate normal distributions (components of the mixture of normal distributions) were used to simulate "true" coefficients for 1000 "customers", where the coefficients were generated for the constant term, contract length, and product discount.

6. Values for fabricated product offers given to the 1000 fabricated customers were randomly generated for contract length and product discount using ranges that approximately matched actual data. The number of offers per customer was generated to match the distribution of actual number of offers per customer in the actual data (1 to 48 offers). Training data included multiple offers per "customer", and test data included one offer for the same customers.

7. The simulated "true" coefficients from step 5 were then used to generate responses to the fabricated offers in step 6. Responses were generated based on a binomial distribution where probabilities were supplied by the binary logistic formula.

The summary statistics for the simulated data sets are reported in Tables 2-5. Note that a contract length of zero years was interpreted as 1 month in the next offer profit (NOP) formula. (The R programs to simulate the fabricated data are available upon request.)

The distribution of number of observations per customer appears in Table 6 below. The distributions of the simulated betas appear in Figures 2, 3, and 4.

Of particular interest is the distribution of the purchase beta. The bimodal distribution indicates a segmented market, with a majority of customers exhibiting strong purchase likelihood, everything else equal. A smaller segment of customers, indicated by the more negative vertical bars at the left side of Figure 2, have substantially lower purchase likelihood.

**Table 2**

**Descriptive Statistics**
**(simulated estimation data)**

|  | id | setnum | X1 | contract length (years) | offer discount | demographic variable (mean centered) | loyalty variable (mean centered) |
|---|---|---|---|---|---|---|---|
| Min. | 1 | 1 | 1 | 0 | -0.49919 | -0.50883 | -0.49794 |
| 1st Qu. | 427 | 1 | 1 | 1 | -0.25898 | -0.23137 | -0.25621 |
| Median | 772 | 1 | 1 | 3 | -0.01007 | 0.003588 | -0.00087 |
| Mean | 671.4405 | 2.748974 | 1 | 2.542522 | -0.00464 | 4.44E-17 | 2.36E-17 |
| 3rd Qu. | 942 | 2 | 1 | 4 | 0.253205 | 0.234932 | 0.257632 |
| Max. | 1000 | 48 | 1 | 5 | 0.499992 | 0.490426 | 0.497247 |
| Number of Observations | 1705 | 1705 | 1705 | 1705 | 1705 | 1000 | 1000 |

**Table 3**
**Dependent Variable Counts**
**(simulated estimation data)**

| No | Yes |
|---|---|
| **665** | **1040** |

**Table 4**

**Descriptive Statistics**
**(simulated test data)**

| | id | setnum | X1 | contract length (years) | offer discount | demographic variable (mean centered) | loyalty variable (mean centered) |
|---|---|---|---|---|---|---|---|
| Min. | 1 | 1 | 1 | 0 | -0.49874 | -0.50883 | -0.49794 |
| 1st Qu. | 250.75 | 1 | 1 | 1 | -0.26147 | -0.23137 | -0.25621 |
| Median | 500.5 | 1 | 1 | 2 | -0.01218 | 0.003588 | -0.00087 |
| Mean | 500.5 | 1 | 1 | 2.444 | -0.00879 | 4.44E-17 | 2.36E-17 |
| 3rd Qu. | 750.25 | 1 | 1 | 4 | 0.24063 | 0.234932 | 0.257632 |
| Max. | 1000 | 1 | 1 | 5 | 0.499361 | 0.490426 | 0.497247 |
| Number of Observations | 1000 | 1000 | 1000 | 1000 | 1000 | 1000 | 1000 |

**Table 5**
**Dependent Variable Counts**
**(simulated test data)**

| No | Yes |
|---|---|
| **373** | **627** |

**Table 6**
**Number of Offer Distribution**
**(simulated estimation data)**

| Number of offers | Percent of customers |
|---|---|
| 1 | 69.0 |
| 2 | 18.1 |
| 3 | 5.5 |
| 4 | 2.7 |
| 5 | 1.7 |
| 6 | 1.3 |
| 7 | .6 |
| 8 | .1 |
| 9 | .3 |
| 10 | .1 |
| 12 | .1 |
| 14 | .1 |
| 15 | .1 |
| 17 | .1 |
| 24 | .1 |
| 48 | .1 |

**Figure 2**

("true" parameter used to simulate estimation and test data)

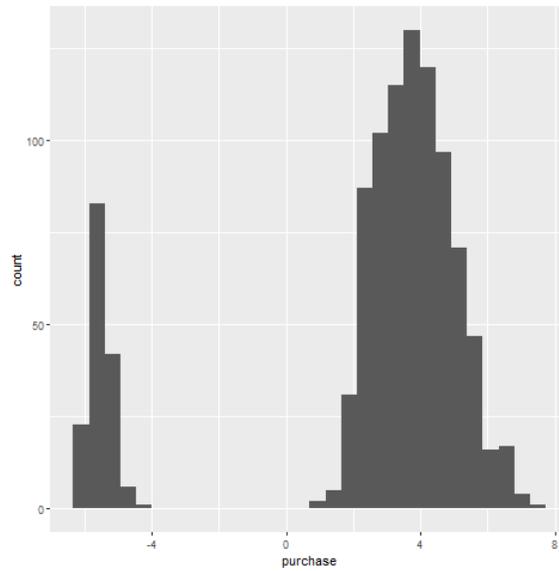

**Figure 3**

("true" parameter used to simulate estimation and test data)

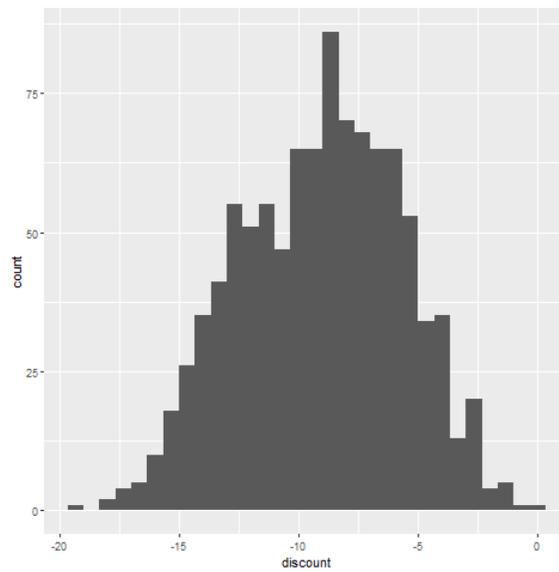

# Figure 4

("true" parameter used to simulate estimation and test data)

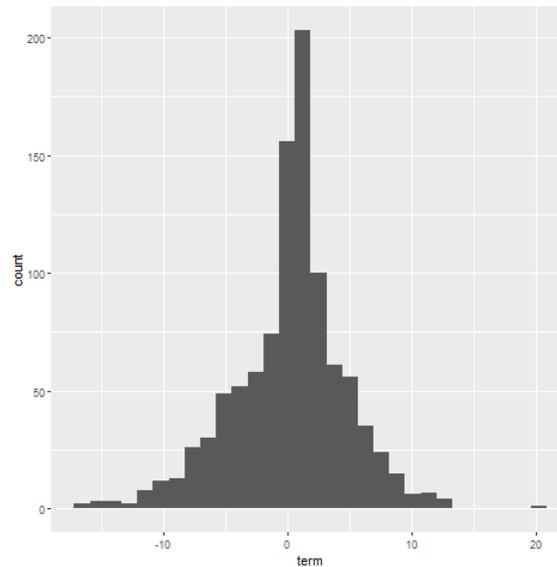

Actual data was used to assess predictive accuracy, since it would not be conclusive to present predictive accuracy results based on simulated data. Publicly available data was acquired that matched as closely as possible the dimensions of actual sales offer data. Unfortunately, no publicly available sales offer data was found that included a history of offers at the customer level. E-commerce sales data were found, however, which differed from the sales offer data in one dimension only, namely, that offers were not proactively given to individual customers.

Just as with the simulated sales offer data, the e-commerce data included:

- Two years (2010 and 2011) of purchases by individual customers.
- Multiple purchase occasions where either a purchase is made or not made.
- B2B transactions (although not exclusively B2B in the case of the e-commerce data).

The purchases were from a UK-based establishment (not a brick-and-mortar store). According the source information, the "company mainly sells unique all-occasion giftware. Many customers of the company are wholesalers" (Dua 2019).

Data was filtered to include purchases of only one product line, gift cups. There were 18 cups in the product line. 2858 purchases were made by 647 customers over approximately 2 years. The data was restructured to include one row for each product on each purchase occasion. In other words, it was assumed that the customer could have chosen any of the 18 products whenever a purchase of any product was made. Multiple products could be purchased on the same occasion. One additional purchase occasion was added for each customer, with a "no purchase" entered for all 18 products. For example, if a customer purchased 9 times, a $10^{th}$ purchase occasion was added for that customer, and there were a total of 180 rows of data for the particular customer. This assumes that the customer could have purchased a cup on a 10$^{th}$ occasion, but decided not to do so.

For each customer, one purchase occasion was selected randomly as validation data, out of all purchase occasions but excluding the first purchase occasion. Then, all remaining purchase occasions were included as training data. This method for data splitting was used to mimic the sales offer database which included multiple purchase occasions for each customer, one for each sales offer, and where the last offer made to each customer would have been selected as validation data.

The split into training and validation data was performed five times, and model hyperparameters were tuned using area under the receiver-operator curve (AUC) averaged across five training-validation splits.

## Model Estimation

The simulated sales-offer data was used to estimate a hierarchical Bayes mixed logit model with a multivariate mixture-of-normals distribution using the R bayesm package (Rossi 2017), explained in detail in chapter 5 of Rossi, et al. (Rossi et al. 2012). The hyperparameter, number of components in the mixture of normals, was tuned using 10 repeats of 10-fold cross-validation resampling using the R caret package (Kuhn et al. 2020a) (R code available upon request). Area under the receiver-operator curve (AUC) was used to pick the optimal value of the hyperparameter. It was found that 1 normal distribution in the mixture-of-normals was optimal. A final model with 1 normal distribution was estimated using the full simulated sales-offer training data.

Using the e-commerce data discussed in the data section of this paper, a hierarchical Bayes multinomial logit model was estimated providing customer-level coefficients. As seen in Table 7, the most accurate model, based on model tuning of the number of mixture-of-normal distributions (ncomp) using 5 repeats of training-validation splits, included 2 normal distributions in the mixture-of-normals and achieved an area under the receiver-operator curve (AUC) of 0.8303 and a predictive accuracy of 0.8685 when predicting validation data (i.e., data not used to estimate the model parameters). The AUC for the hierarchical Bayes model with test data was 0.8430 (Table 9).

AUC was used as the measure of model performance for the purpose of tuning ncomp. Unlike predictive accuracy (third column in Table 7), which uses a particular probability threshold to label purchase as yes (if probability>percent purchase in training data, then predict that product is purchased), AUC does not assume a particular threshold. Instead, AUC incorporates all possible probability thresholds to measure performance and avoids the possibility of multiple solutions for the optimal value of ncomp.

**Table 7**

| Hierarchical Bayes Mixed-Logit Results of E-Commerce Data |||
| :---: | :---: | :---: |
| (5 Repeats of Training-Validation Resampling) |||
| Number of Components of Mixture-Of-Normal Distribution (ncomp) | Area Under Receiver Operator Curve (AUC) | Percent of Test Data Observations Correctly Predicted |
| 1 | 0.8300 | -- |
| 2 | 0.8303 | 0.8685 |
| 3 | 0.8199 | -- |
| 4 | 0.8237 | -- |
| 5 | 0.8173 | -- |

## Results – Predictive Accuracy

For the random forest model (Table 8), the maximum AUC achieved was 0.9160 and a predictive accuracy of 0.8473. The AUC for the random forest model with test data was 0.9156 (Table 9).

## Table 8

| Random Forest Results for E-Commerce Data (5 Repeats of Training-Validation Resampling) | | |
|---|---|---|
| Number of Randomly- Drawn Variables Per Tree (mtry) | Area Under Receiver Operator Curve (AUC) | Percent of Test Data Observations Correctly Predicted |
| 2 | 0.8548 | -- |
| 10 | 0.9127 | -- |
| 21 | 0.9160 | 0.8473 |

## Table 9

| Area Under Receiver Operator Curve (AUC) (based on e-commerce test data) | |
|---|---|
| Model | AUC |
| Mixed Logit | 0.8430 |
| Random Forest | 0.9156 |

Based on validation data (Tables 7 and 8), the hierarchical Bayes model predicted 1.74 times better than a chance model (0.8685/0.5). It predicted offer acceptance less accurately than a random forest model based on both predictive accuracy (0.8685 vs. 0.8473) and AUC (0.8303 vs. 0.9160).

After splitting the e-commerce data one final time into training and test and re-estimating the two models, HB mixed logit and random forest with their optimal hyperparameter values, a Delong, Delong, and Clarke-Pearson test (DeLong et al. 1988) statistic was used to test the null hypothesis that there is no difference between the two receiver-operator curves (mixed logit vs. random forest).

The Delong, Delong, and Clarke-Pearson statistic had a p-value 6.203e-10; thus, the null hypothesis of no difference between the performance of the two models (mixed logit vs. random forest) was rejected.

The e-commerce results suggest that HB mixed logit with customer-level parameter estimates does not predict new data as accurately as random forest. However, the AUC score for HB mixed logit was strong (0.830 versus 0.916 for random forest), and the lift for the HB mixed logit model (Figure 5) was impressive. The lift curve indicates that the top 20% with the highest predicted purchase probability would successfully identify almost 80% of actual purchasers.

Figure 5

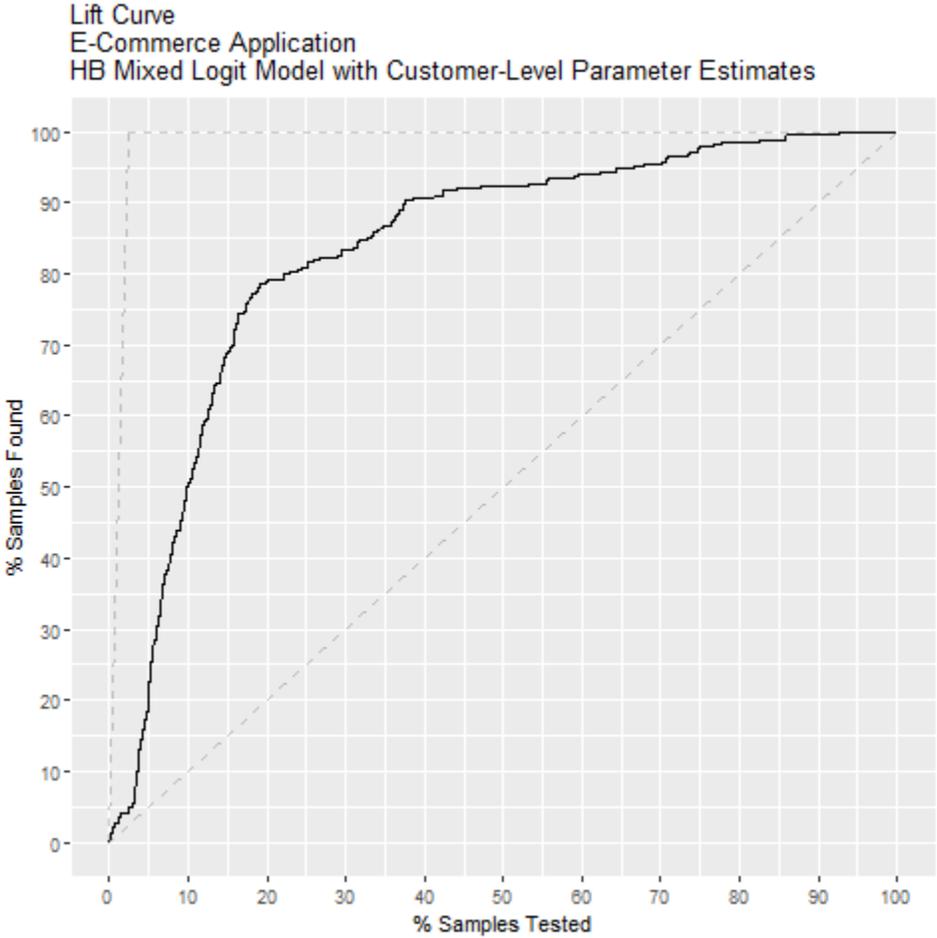

The HB estimation procedure produces individual customer-level parameter estimates, and the test data included observations for new data (purchase occasions not in the training data) for the same 647 customers as were in the training data.

The relatively strong predictive performance for HB mixed logit may be due to the use of customer-level parameters estimates to predict the test data. Furthermore, the feature of HB modeling, that customers with only one purchase occasion in the sales database borrow information from customers with multiple offers, likely produced a boost in predictive accuracy.

## Results – Optimization

The customer-level parameter estimates of the mixed logit model were used together with simulated test data, another set simulated offers (offers not in the training data) for the same 1000 simulated customers as were in the training data to maximize NOP.

Upper and lower bounds of 50% and -50% were imposed on the discount rates, and contract length was restricted to 5 options (1 month, 12 months, 24 months, 36 months, and 60 months). Using the Python pyomo package and the Bonmin nonlinear mixed-integer programming software, the following NOP-maximized results were found (Table 10).

**Table 10**
**Optimal Discount Rate ($r$) and Contract Length ($m$)**

|  | Not Loyal | Loyal |
|---|---|---|
| **Discount Inelastic** | $r = +50.0\%$ <br><br> $m = 60$ months | $r = +50.0\%$ <br><br> $m = 60$ months |
| **Discount Elastic** | $r = -30.8\%$ <br><br> $m = 12$ months | $r = -41.8\%$ <br><br> $m = 24$ months |

The results suggest that the lowest discount and the longest contract term, a 50.0% premium and 60 months, respectively, should be charged to discount inelastic customers, regardless whether they are loyal or not. Few enough of these customers would churn (discontinue as a customer) with a higher unit price, so that locking them into a 60-month contract becomes a more profitable long-term strategy.

In contrast, discount elastic customers should be offered substantial discounts and shorter contract lengths, in order to entice them into a contract. Among these discount elastic customers, the results suggest higher discounting with a longer contract, 42% vs. 31% and 24 months vs. 12 months, for loyal vs. non-loyal customers, respectively. This outcome might describe a situation where a customer has been historically loyal but is sensitive to internal pressure to reduce costs, making a high discount required. Customers who are discount elastic and not loyal can only be captured for 12 months, albeit with a lesser discount.

These results could and probably would be very different for different companies, depending on the strength of the brand, competitive pressure, and customers' need for cost reduction.

As the optimal solution requires targeting the discount elastic customer segment, one would wonder how well the model would find those customers who are within this segment. Figure 6 shows the lift curve for these customers, based on simulated test data.

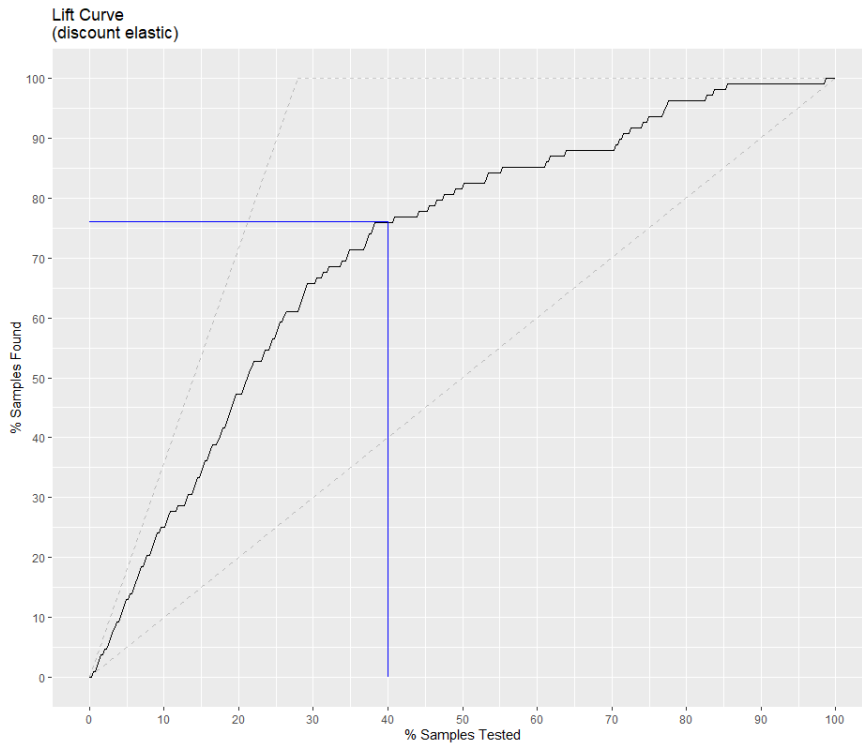

**Figure 6**
**(based on simulated test data)**

Recall that the discount elastic customers were identified based on the model itself (38.7% of customers within the offer database, according to Table 1). If the model were then used to target the most likely purchasers among these 38.7%, the lift curve indicates that the top 40% with the highest predicted purchase probability would successfully identify more than 75% of actual purchasers.

## Discussion

Among businesses competing in a business-to-business market with strong competition, many companies feel compelled to increase discounts each year in order to retain customers. This paper demonstrates end-to-end analysis using a customer offer database to estimate customer-level choice models and applies those choice models to build an optimization tool able to support value-based pricing and selling. In particular, discount pricing strategy is addressed within the context of next-offer profits and customer loyalty.

From a technical standpoint, this case study provides an example of combining machine learning, hierarchical Bayes choice modeling, and nonlinear mixed-integer programming to provide an analytical tool to support rapid investigation of sales strategies. (The nonlinear mixed-integer programming algorithm solved in about 1 minute on a typical personal computer.) It should also be noted that in other applications of this method with actual in-market data, Bonmin was found to produce the same results as produced by AMPL, a commercial software that combines best-in-class solvers—BARON, Gurobi, and Knitro (AMPL 2020)—suggesting that the use of open-source software for this type of optimization problem will produce accurate optimizations.

It is worth noting that the results from Table 10 represent viable strategies for a company's sales team in a Business-to-Business setting in which discounts can be different for each customer (or group of customers).

While only three dimensions (discounting, contract length, and loyalty) of sales strategy were addressed in this study, the methodology can be extended to explore other strategic business decisions, such as cross-subsidization of product lines. Similarly, the methodology can easily be extended to incorporate more complex cost functions into the NOP objective function.

The findings from this research suggest that:

1) A sales offer database in a B2B technology market can provide sufficient variability in price and features to estimate a usable mixed logit model with customer-level parameter estimates.
2) Mixed logit predictive accuracy compares favorably with that of random forest, a machine learning model that consistently provides high predictive accuracy. While the predictive accuracy results cannot be considered conclusive for sales offer data, results for e-commerce data with similar dimensions (number of months of purchase history and number of observations per customer) suggests that predictive accuracy with customer-level mixed logit parameter estimates is strong, as it compares favorably to random forest.
3) Mixed logit parameter estimates can be incorporated into a next-offer profit maximization algorithm which solves rapidly, suggesting that the methodology used in this study could provide a valuable tool for value-based pricing and selling efforts in the B2B sector.

## References

Aboutaleb, Youssef M, et al. (2021), 'Discrete Choice Analysis with Machine Learning Capabilities', *arXiv preprint arXiv:2101.10261*.
Acuna-Agost, Rodrigo, Thomas, Eoin, and Lhéritier, Alix (2021), 'Price elasticity estimation for deep learning-based choice models: an application to air itinerary choices', *Journal of Revenue and Pricing Management*, 1-14.
Agarwal, James, et al. (2015), 'An interdisciplinary review of research in conjoint analysis: recent developments and directions for future research', *Customer Needs and Solutions,* 2 (1), 19-40.
AMPL (2020), 'Solvers We Sell', *AMPL* <https://ampl.com/products/solvers/solvers-we-sell/>, accessed.

Ben-Akiva, Moshe E. and Lerman, Steven R. (1985), *Discrete choice analysis : theory and application to travel demand* (MIT Press series in transportation studies; Cambridge, Mass.: MIT Press) xx, 390 p.
Breidert, Christoph, Hahsler, Michael, and Reutterer, Thomas (2006), 'A review of methods for measuring willingness-to-pay', *Innovative Marketing,* 2 (4), 8-32.
Classen, Moritz and Friedli, Thomas (2019), 'Value-Based Marketing and Sales of Industrial Services: A systematic literature review in the age of digital technologies', *Procedia Cirp,* 83, 1-7.
DeLong, Elizabeth R., DeLong, David M., and Clarke-Pearson, Daniel L. (1988), 'Comparing the areas under two or more correlated receiver operating characteristic curves: a nonparametric approach', *Biometrics*, 837-45.
Dua, Dheeru and Graff, Casey 'Online Retail II Data Set', <https://archive.ics.uci.edu/ml/datasets/Online+Retail+II>, accessed.
Elrod, Terry, Louviere, Jordan J., and Davey, Krishnakumar S. (1992), 'An empirical comparison of ratings-based and choice-based conjoint models', *Journal of Marketing research,* 29 (3), 368-77.
Fader, Peter S. and Hardie, Bruce G. S. (2016), 'Reconciling and clarifying CLV formulas'.
Feldman, Jacob, et al. (2018), 'Customer choice models versus machine learning: Finding optimal product displays on Alibaba', *Available at SSRN 3232059*.
Gourieroux, Christian and Monfort, Alain (1993), 'Simulation-based inference: A survey with special reference to panel data models', *Journal of Econometrics,* 59 (1-2), 5-33.
Hart, William E., et al. (2017), *Pyomo-optimization modeling in python* (67: Springer).
Hastie, Trevor, Tibshirani, Robert, and Friedman, Jerome (2009), *The elements of statistical learning: data mining, inference, and prediction* (Springer Science & Business Media).
Hensher, David, Shore, Nina, and Train, Kenneth (2005), 'Households' willingness to pay for water service attributes', *Environmental and Resource Economics,* 32 (4), 509-31.
Hensher, David A. and Greene, William H. (2003), 'The mixed logit model: the state of practice', *Transportation,* 30 (2), 133-76.
Hess, S. and Train, K. (2017), 'Correlation and scale in mixed logit models. J. Choice Model. 23, 1–8'.
Hinterhuber, Andreas (2004), 'Towards value-based pricing—An integrative framework for decision making', *Industrial marketing management,* 33 (8), 765-78.
--- (2008), 'Value delivery and value-based pricing in industrial markets', *Creating and managing superior customer value* (Emerald Group Publishing Limited).
--- (2017), 'Value quantification capabilities in industrial markets', *Journal of Business Research,* 76, 163-78.
Hinterhuber, Andreas and Liozu, Stephan M. (2018), 'Thoughts: premium pricing in B2C and B2B', *Journal of Revenue and Pricing Management,* 17 (4), 301-05.
Hinterhuber, Andreas, Snelgrove, Todd C., and Stensson, Bo-Inge (2021), 'Value first, then price: the new paradigm of B2B buying and selling', *Journal of Revenue and Pricing Management*, 1-7.
Huber, Joel and Train, Kenneth (2001), 'On the similarity of classical and Bayesian estimates of individual mean partworths', *Marketing Letters,* 12 (3), 259-69.
Johnson, Richard M. (2000), 'Understanding HB: an intuitive approach', *Sawtooth Software Inc., Sequim, WA.*

Kamakura, Wagner A. and Russell, Gary J. (1989), 'A probabilistic choice model for market segmentation and elasticity structure', *Journal of marketing research,* 26 (4), 379-90.
Kuhn, Max (2019), 'caret: Classification and Regression Training', (R package version 6.0-84; https://CRAN.R-project.org/package=caret).
Kuhn, Max, et al. (2020a), 'Package 'caret'', *The R Journal*.
--- (2020b), 'Package 'caret'', *The R Journal*, 223.
Lhéritier, Alix, et al. (2019), 'Airline itinerary choice modeling using machine learning', *Journal of choice modelling,* 31, 198-209.
Li, Zheng, Hensher, David A., and Rose, John M. (2010), 'Willingness to pay for travel time reliability in passenger transport: A review and some new empirical evidence', *Transportation research part E: logistics and transportation review,* 46 (3), 384-403.
Liaw, Andy and Wiener, Matthew (2002), 'Classification and regression by randomForest', *R news,* 2 (3), 18-22.
Lu, Jing, et al. (2021), 'Modeling hesitancy in airport choice: A comparison of discrete choice and machine learning methods', *Transportation Research Part A: Policy and Practice,* 147, 230-50.
Masiero, Lorenzo, Heo, Cindy Yoonjoung, and Pan, Bing (2015), 'Determining guests' willingness to pay for hotel room attributes with a discrete choice model', *International Journal of Hospitality Management,* 49, 117-24.
McFadden, Daniel and Train, Kenneth (2000), 'Mixed MNL models for discrete response', *Journal of applied Econometrics,* 15 (5), 447-70.
Milte, Rachel, et al. (2018), 'What characteristics of nursing homes are most valued by consumers? A discrete choice experiment with residents and family members', *Value in Health,* 21 (7), 843-49.
Newman, Jeffrey P., et al. (2014), 'Estimation of choice-based models using sales data from a single firm', *Manufacturing & Service Operations Management,* 16 (2), 184-97.
Orme, Bryan and Howell, John (2009), 'Application of covariates within Sawtooth Software's CBC/HB program: theory and practical example', *Sawtooth Software Research Paper Series*.
Paneque, Meritxell Pacheco, et al. (2021), 'Integrating advanced discrete choice models in mixed integer linear optimization', *Transportation Research Part B: Methodological,* 146, 26-49.
Pedregosa, Fabian, et al. (2011), 'Scikit-learn: Machine learning in Python', *the Journal of machine Learning research,* 12, 2825-30.
Pietro Belotti, Pierre Bonami, Claudia D'Ambrosio, John J. Forrest, Laszlo Ladanyi, Carl Ladanyi, Carl Laird, John Lee, Francois Margot, Stefan Vigerske, Andrew Waechter (2019), 'Basic Open-source Nonlinear Mixed INteger programming', (1.8.7 edn.; https://github.com/coin-or/Bonmin/wiki).
Pöyry, Essi, Parvinen, Petri, and Martens, Jonas (2021), 'Effectiveness of value calculators in B2B sales work–Challenges at the sales-call level', *Journal of Business Research,* 126, 350-60.
Ratliff, Richard M., et al. (2008), 'A multi-flight recapture heuristic for estimating unconstrained demand from airline bookings', *Journal of Revenue and Pricing Management,* 7 (2), 153-71.
Regier, Dean A., et al. (2009), 'Bayesian and classical estimation of mixed logit: an application to genetic testing', *Journal of health economics,* 28 (3), 598-610.

Rossi, Peter E. (2017), 'bayesm: Bayesian Inference for Marketing/Micro-Econometrics', (R package version 3.1-0.1; https://CRAN.R-project.org/package=bayesm).
Rossi, Peter E., Allenby, Greg M., and McCulloch, Rob (2012), *Bayesian statistics and marketing* (John Wiley & Sons).
Saito, Taiga, et al. (2019), 'Application of online booking data to hotel revenue management', *International Journal of Information Management,* 46, 37-53.
Sonnier, Garrett, Ainslie, Andrew, and Otter, Thomas (2007), 'Heterogeneity distributions of willingness-to-pay in choice models', *Quantitative Marketing and Economics,* 5 (3), 313-31.
Talluri, Kalyan and Van Ryzin, Garrett (2004), 'Revenue management under a general discrete choice model of consumer behavior', *Management Science,* 50 (1), 15-33.
Töytäri, Pekka and Rajala, Risto (2015), 'Value-based selling: An organizational capability perspective', *Industrial Marketing Management,* 45, 101-12.
Train, Kenneth (2001), 'A comparison of hierarchical Bayes and maximum simulated likelihood for mixed logit', *University of California, Berkeley*, 1-13.
Train, Kenneth and Weeks, Melvyn (2005), 'Discrete choice models in preference space and willingness-to-pay space', *Applications of simulation methods in environmental and resource economics* (Springer), 1-16.
Train, Kenneth E. (2009), *Discrete choice methods with simulation* (Cambridge university press).
Van Cranenburgh, S., et al. (2021), 'Choice modelling in the age of machine learning', *arXiv preprint arXiv:2101.11948*.
Walker, Joan and Ben-Akiva, Moshe (2002), 'Generalized random utility model', *Mathematical social sciences,* 43 (3), 303-43.
Wardell, Clarence L., Wynter, Laura, and Helander, Mary (2008), 'Capacity and value based pricing model for professional services', *Journal of revenue and pricing management,* 7 (4), 326-40.
Wittink, Dick R. and Cattin, Philippe (1989), 'Commercial use of conjoint analysis: An update', *Journal of marketing,* 53 (3), 91-96.
Wu, Caesar, Buyya, Rajkumar, and Ramamohanarao, Kotagiri (2019), 'Value-based cloud price modeling for segmented business to business market', *Future Generation Computer Systems,* 101, 502-23.
Zhao, Xilei, et al. (2020), 'Prediction and behavioral analysis of travel mode choice: A comparison of machine learning and logit models', *Travel behaviour and society,* 20, 22-35.